\documentclass{article}
\usepackage{amsmath}
\usepackage{hyperref}

\title{Cut-Paste Metrics: A More Natural Approach to Approximate String Matching}
\author{Paul Burchard}
\date{1995}

\newcommand{\Paste}{\mathop{\text{Paste}}}
\newcommand{\Insert}{\mathop{\text{Insert}}}

\newcommand{\kgoal}{k_{\text{goal}}}

\begin{document}
\maketitle

\begin{abstract}
Despite the breadth of application of approximate string matching, the current body of theoretical work is largely based on a metric (the so-called {\em edit metric\/}) which does not qualitatively reflect the principal processes through which strings change in applications.  To remedy this, we introduce the notion of {\em cut-paste metrics}, in which substring reordering can be accomplished at sublinear cost.

We then exhibit a specific cut-paste metric for which best matches can be practically computed.  With a text string of length $n$, a search string of length $m$, and at most $k$ paste and insertion errors allowed, the text substring closest to the search string can be found in $O(n k\log m)$ time and $O(m(k+\log m))$ space, provided that the search string has a suffix tree of depth $O(\log m)$ (which is typical).

We furthermore demonstrate a non-symmetric cut-paste metric which allows best matches containing long gaps to be computed in the same time and space.
\end{abstract}

\section{Introduction}

Approximate string matching is an important problem in a variety of fields, including genome informatics and natural language databases.  When the meaning of ``approximate'' is specified in more detail, however, tradeoffs must always be made between speed and sophistication, and between generality and domain-specificity.  For example, a human searching through a large database of natural language documents is extremely sophisticated, but at the same time slow, expensive, and highly specialized for this sort of task.

We believe that the tradeoff made by the standard {\em edit metric\/} is not adequate for many applications, where the natural processes of change imposed on strings involve more than just the single-letter ``mutations.''  Substring reordering---``cutting and pasting''---is another key process, whose effects are captured only when the metric allows substring reordering to be performed at {\em sublinear\/} cost relative to the equivalent insertions and deletions.  We will exhibit a computationally practical metric with this {\em cut-paste\/} property.

A related process, which comes from reordering substrings in a larger string context, is the insertion of long foreign substrings, or ``gaps.''  This effect has been incorporated in previous work based on the edit metric (see e.g.~\cite{bib.goto.longgap}).  We will offer a unified treatment of all three effects by exhibiting a closely related non-symmetric distance function which has practically computable best matches, and assigns sublinear cost to ``gaps'' in the text string, while retaining their usual linear cost in the search string.

The most obvious cut-paste metric would be one that counts the minimum number of iterated cut-paste and single-letter-insert operations required to transform one string into another.  However, this does not appear to be a computationally practical approach.  For this reason, we define the {\em copy-paste metric\/} as the logarithm of the minimum number of one-pass (non-overlapping, non-iterated) copy-paste and single-letter-insert operations sufficient to assemble either string from the other one.

The combinatorics of the copy-paste metric are simple enough to permit the use of efficient {\em greedy algorithms\/} for locating best matches.  At the same time, the (unavoidable) logarithm in the definition is a vestigal reminder that iteration of copy-paste operations can have exponential effects.  Aside from the combinatorics, the main task for the best-match algorithm is the {\em dynamic inversion\/} of substring matching statistics.  This is accomplished through careful use of the search string's suffix tree as a cache, and results in performance within a $\log(m)$ factor of algorithms based on the edit metric ($m$ being the length of the search string).

As a concrete example of the foregoing, consider a pair of strings taken from drafts of this paper:
\begin{verbatim}
   With at most $k$ paste and insert errors allowed, a search
   string of length $m$, and a text string of length $n$

   the text string, $m$ the length of the search string, and
   $k$ the number of paste and insert errors allowed
\end{verbatim}
The copy-paste distance between these strings is $\log_2(1+23)$, representing only a 21\% discrete error rate ($23/107 \approx 0.21$).  The edit metric, however, would consider these two strings to be essentially unrelated.

\section{Cut-Paste String Metrics}

Choose a finite alphabet $A$.  An {\em indexed string\/} is a map $I\mathop\rightarrow\limits^X A$ where $I$ is a finite ordered set.  An {\em indexed substring\/} of $X$ is a composition $J \mathop\rightarrow\limits^f I \mathop\rightarrow\limits^X A$ where $J \mathop\rightarrow\limits^f I$ is injective.  The broader term {\em substring\/} includes any indexed string which may be made into an indexed substring by some injective map of indices.

Suppose we wish to produce the indexed string $J \mathop\rightarrow\limits^Y A$ by performing editing operations on a source string $I \mathop\rightarrow\limits^X A$.  The transformation resulting from the editing operations can be encoded in a partially-defined map of indices $J \mathop{\cdots\rightarrow}\limits^f I$ in the reverse direction, indicating where (if anywhere) each letter of $Y$ originated in $X$.  We of course require that $Y(j) = X(f(j))$ whenever $f$ is defined.

Each such an index correspondence $f$ can be generated by the iteration of elementary editing operations in many ways.  But there is a unique simplest {\em one-pass\/} method for generating $f$ in terms of Paste and Insert operations.  To be precise, the operations allowed are {\em One-pass Paste\/}: appending a (contiguous) substring of $X$; and {\em One-pass Insert\/}: appending a single letter of $A$.  The number of one-pass paste and insert operations in the minimal decomposition of the index correspondence are, respectively,
\begin{eqnarray*}
   \Paste(f) &:=& \# \{ j \mathrel{|} \exists f(j) \ \wedge\ \neg\  f(j-1) = f(j)-1 \},\\
   \Insert(f) &:=& \# \{ j \mathrel{|} \not\exists f(j) \}.
\end{eqnarray*}

To build a metric out of these numbers, we choose a weight $w > 0$, which specifies the number of Insert operations equivalent in cost to a single Paste; and a convex homogeneous order-preserving symmetric real-valued function $\rho$ on the cone $[0,\infty) \times [0,\infty)$, positive except at the origin, which will be used to symmetrize the metric.  We will later specialize to the case $w=1$, $\rho=\max$ in order to get good combinatorics (although to its detriment, $\rho=\max$ will produce a larger set of ``best matches'').

The {\em one-sided copy-paste distance\/} $\delta(X,Y)$ is then defined to be the binary logarithm of the minimum weighted number of paste and insert operations which, when copying from $X$, will produce $Y$ in one pass:
\[
   \delta(X,Y) :=
   \min_{\def\arraystretch{0.7}\begin{array}{c}
      \scriptstyle f : J \cdots\rightarrow I\\
      \scriptstyle Y(j) = X(f(j))\end{array}}
   \log_2 \left(\strut \Paste(f) + \tfrac{1}{w} \Insert(f) \right).
\]
The {\em copy-paste metric\/} $d(X,Y)$ is then obtained from $\delta(X,Y)$ by symmetrization:
\[
   d(X,Y) := \rho(\delta(X,Y),\,\delta(Y,X)).
\]

\paragraph{Lemma 1.}
For any $w$ and $\rho$ as above, $d(X,Y)$ is a metric on the set of strings of length $> w$.

\paragraph{Proof.}
The main issue is the triangle inequality.  Given two index correspondences $K \mathop{\cdots\rightarrow}\limits^g J$ and $J \mathop{\cdots\rightarrow}\limits^f I$, composition yields
\begin{eqnarray*}
   \Paste(f\circ g) &\le& \Paste(f) \Paste(g), \\
   \Insert(f\circ g) &\le& \Insert(f) \Paste(g) + \Insert(g).
\end{eqnarray*}
If $\# J \ge w$, then $\Paste(f) + \tfrac{1}{w}\Insert(f) \ge 1$ for every index correspondence $f$.  Combining this with the above inequalities, we have
\begin{eqnarray*}
   &&\Paste(f\circ g) + \tfrac{1}{w}\Insert(f\circ g) \le {} \\
   &&\qquad (\Paste(f) + \tfrac{1}{w}\Insert(f))
   (\Paste(g) + \tfrac{1}{w}\Insert(g)).
\end{eqnarray*}
Applying a logarithm to both sides now yields the triangle inequality for $\delta(X,Y)$.  The same then holds for $d(X,Y)$, because any symmetrization function $\rho$ with required properties preserves the triangle inequality.

In order to handle the ``gap'' effect, we must abandon the symmetry requirement in our measure of distance---after all, we do not want to allow long non-matching substrings to occur in the search pattern, only in the text.  On the other hand, $\delta(X,Y)$ is not useful by itself, since for fixed $Y$ it is always minimized by maximizing $X$.  We need to impose {\em some\/} cost on extraneous letters in $X$.

We will therefore ``partially symmetrize'' $\delta$ by defining a looser ``gap'' version $\delta_{\ell}$ of $\delta$ and then using the non-symmetric metric $d_{\ell}(X,Y) := \max(\delta(X,Y),\,\delta_{\ell}(Y,X))$.  The distance function $\delta_{\ell}$ will not necessarily satisfy the triangle inequality on its own, but will obey $\delta_{\ell}(Z,X) \le \delta_{\ell}(Z,Y) + \delta(X,Y)$, enough to make $d_{\ell}(X,Y)$ a non-symmetric metric.  Moreover, $\delta_{\ell}$ will reduce to the original $\delta$ when $\ell = 1$.

[---under construction---]

\section{Algorithm for Best Matching Substring}

In the description of the algorithm, we will adopt some terminology from~\cite{bib.chla.matchbio}.  The {\em substring matching statistics\/} of one string relative to another constitute a function computing the longest common substring of the two with a given starting location in the first.  The term {\em jump\/} (relative to another string) refers to motion from one location in a string, to that location just past the end of the longest common substring.  Following the standard picture from dynamic programming, we will think of the text string as {\em horizontal\/} and the search string as {\em vertical}.  Thus, {\em vertical jumps\/} will refer to motion along the search string (according to its longest common substrings with the current candidate substring of the text).

As mentioned in the introduction, there are two key points in the algorithm.  First, substring matching statistics allow us to easily identify the optimal way to paste together one string out of pieces of another.  That's because---at least when $w \le 1$---optimization can be achieved by a {\em greedy algorithm\/} which simply follows successive jumps.  Second, although the techniques of~\cite{bib.chla.matchbio} allow efficient generation of matching statistics of the text string relative to the search string, careful caching of these statistics inside the suffix tree is required in order to efficiently calculate the inverse matching statistics of the search string relative to the current search window of the text.

\subsection{Data Structures.}

For each search string, we prepare an enhanced version of {\em suffix tree\/} data structure of~\cite{bib.mccr.sufftree} and~\cite{bib.chla.matchbio}.  In the basic tree, each tree node represents a range of substrings of the given string, maximal with respect to the property that any indexed realization of one substring implies indexed realizations of all the others starting at the same location in the given string.  The leaves correspond to the suffixes of the given string, whence the name.  We will describe the necessary enhancements below.

During each search of a text string, we maintain a {\em search window\/} data structure consisting of two back-to-back bidirectional {\em match queues\/} and a copy of the search string's enhanced suffix tree.  The queue entries correspond to the text locations inside the current search window, and are split between the {\em full match queue\/} on the left, and the {\em partial match queue\/} on the right.  The partial match queue only holds an entry as long as the longest exact match at the corresponding text location extends to at least the right end of the search window, while the full match queue holds an entry for the rest of its sojourn in the search window (when the longest exact match at the corresponding text location is fully contained within the search window).  Each queue entry consists of a text location, a pointer into the suffix tree, and a truncation value indicating which of the substrings corresponding to that suffix tree node is the longest common substring at that text location.

The purpose of the two queues is to maintain ``reference counts'' in the suffix tree, in order to support rapid dynamic inversion of substring matching statistics---that is, the computation of the matching statistics of the search string relative to the text substring inside the search window.  The enhancements to the suffix tree required to perform this computation are:
\begin{itemize}
\item
A fixed global array indexing the leaves of the suffix tree by search string location.
\item
A fixed per-node pointer caching the {\em shift\/} operator (left truncation), as already included in~\cite{bib.mccr.sufftree} and~\cite{bib.chla.matchbio}.
\item
A dynamic per-node sparse sorted list of reference counts for the possible truncations of that node (sorted by truncation).
\item
A dynamic per-node integer reference count for the non-truncated occurrences of that node.
\item
A dynamic per-node partial match queue for entries of the global partial match queue which correspond to that node or its children.
\end{itemize}
The first two enhancements may be prepared in advance, per search string, without time or space penalty.  The last three enhancements are critical to the efficient computation of inverse matching statistics.  The sparse sorted lists may be implemented either deterministically as balanced trees, or probabilistically as skip lists~\cite{bib.pugh.skiplist}; we assume that the minimum truncation with non-zero reference count can be looked up in constant time.

\subsection{Invariants.}

The {\em jump length\/} of the combined match queue is $\kgoal\le k$, and its actual length is never more than $m \kgoal$.  The full match portion has jump length $\kgoal - 1$, and the partial match portion has jump length $1$.  $\kgoal$ is a non-increasing function during the run of the algorithm.

The search window's enhanced suffix tree contains reference counts of all the substrings represented by entries in the full match queue, and the (non-contiguous) subqueue of the partial match queue consisting of all entries pointing to that node.

\subsection{Algorithm.}

\begin{description}

\item[1. Per-Search-String Preparation.]
Build the suffix tree of the search string, with shift pointers and leaf index, as in~\cite{bib.mccr.sufftree} or~\cite{bib.chla.matchbio}.

\item[2. Per-Search Preparation.]
Set the search window to an empty window at the beginning of the text string.  Its enhanced suffix tree should be a copy of the enhanced suffix tree of the search string, with all truncation refcount lists and partial match queues empty.

Set $\kgoal = k$.

\item[3. Search Loop.]
For each position in the text string, the following operations are performed:
\begin{description}

\item[A. Advance Right Edge of Window.]
Run Chang and Lawler's on-line algorithm~\cite[Sec.~2.2]{bib.chla.matchbio} for substring matching statistics until the combined match queue been generated to jump length exactly $\kgoal$ from the left edge of the window (or has hit the end of the text string), with exactly one jump remaining in the partial match part.  Put each new entry into the appropriate node-specific partial match queue, and then when advancing the full/partial match boundary, take each entry off the node-specific partial match queue and increment the corresponding entry in that node's refcount list. 

\item[B. Perform Vertical Probe.]
Starting at the bottom end of the search string, advance at most $\kgoal$ jumps.  These longest matches are computed by going up the suffix tree from the leaf node corresponding to the current position in the search string, until we find a nonempty refcount list.  We also keep track of the minimum text position in the node-specific partial match queues we encounter along the way.  The longest match corresponds either to the smallest truncation value in the first nonempty refcount list, or the minimum starting position of partial match entries encountered up to then (which can be translated into a truncation value for that node based on the current left-hand boundary of the partial match queue).

\item[C. Make Window Narrower If Successful.]
If the number of vertical jumps required to cover the search string is found to be less than or equal to $\kgoal$, we now have a new best match.  Record the match, and set $\kgoal$ to one less than the larger of the horizontal and vertical jump counts.  Retract the right edge of the search window, if necessary, so that it again covers exactly $\kgoal$ horizontal jumps from the left edge of the window.  Remove corresponding refcounts, and push and pop the corresponding node-specific partial match queue entries.

If the number of vertical jumps needed was actually less than or equal to the {\em new\/} $\kgoal$, go back to Step 3.B.{} now.

\item[D. Advance Left Edge of Window.]
Move the left edge of the search window one unit forward, deleting the leftmost node of the match graph, and decrementing the corresponding refcount in the suffix tree cache.

\end{description}

\end{description}

\subsection{Analysis.}

\paragraph{Theorem 1.}
Consider the copy-paste string metric with $w=1$, $\rho=\max$.  Let $n$ be the length of the text string, $m$ the length of the search string, $h$ the depth of the search string's suffix tree, and $k$ the maximum number of paste and insert errors to be permitted.  Assume that $m = o(n)$, $k = O(m)$, $k>0$.  Then the above algorithm correctly locates one text substring which best matches the search string, consuming $O(n(\log m + kh))$ time and $O(m(h+k))$ space in the process.

\paragraph{Correctness.}
First, we verify that a greedy algorithm applied to substring matching statistics give the longest substring coverable with $\le k$ errors, as long as $w=1$.  [If we catch up with an advanced wave by skipping a letter, it's no gain.]

As a subsidiary point, note that $\rho=\max$ is required so that the greedy algorithms along the two strings can be run ``lexicographically.''  Use of $\max$ ensures that we encounter exactly one minimum distance in such a lexicographic procedure.

Second, we verify that our caching correctly inverts the matching statistics.  Given a match in the text string, how do we ensure that all of its substrings in the search string can be found?  Recall that we travel up the suffix tree, checking all the right-truncated substrings at a search string location.  Right truncation of the text match is is handled by the propagation of the cache information up the tree, where it can find common right-truncated ancestors of the suffix leaf being searched.  Left truncation of right truncates is handled automatically by the queue---all ``shifts'' of a text match which intersect the search window are in the queue, and will be cached in the same way as the text match itself.

\paragraph{Time.}
At most $k$ narrowings of the window will be performed over the whole run, so since $k=o(n)$ we can assume that the window boundaries only move forward.  We can also ignore the horizontal matching statistics, since we are computing them on-line at a cost of only $O(n)$ time and $O(m)$ space.

At each of $n$ text locations, then, we will perform $\le k$ horizontal jumps, $\le k$ vertical jumps, and one increment of the left, middle, and right edges of the search window past that location.
\begin{itemize}
\item
Each horizontal jump takes constant time, given the horizontal matching statistics.
\item
Each vertical jump takes $O(h)$ time, since the caching has been designed to perform inversion of matching statistics by traveling up the suffix tree at constant lookup cost per suffix node encountered.
\item
Each increment of the left edge of the search window takes $O(\log m + h)$ time, corresponding to one injection into one node's sparse list of truncation refcounts (of virtual length $\le m$), plus $\le h$ increments of non-truncation refcounts in its ancestors.
\item
Each increment of the right edge of the search window takes $O(h)$ time, corresponding to $\le h$ injections into node-specific partial match queues.
\item
Each increment of the middle edge of the search window is equivalent to the combination of a left- and a right-edge increment.
\end{itemize}
Thus, the total time is $O(n(kh + k + \log m + h) = O(n(\log m + kh))$ as claimed.

\paragraph{Space.}
The storage consists of the search window queues, the enhanced suffix tree with caching, and lists of horizontal and vertical jumps.
\begin{itemize}
\item
The search window takes $O(m k)$ storage, queueing up horizontal matching statistics for $\le m k$ text positions.
\item
The enhanced suffix tree with caching takes $O(m(h+k))$ storage.  As is known, the tree has $O(m)$ nodes, and requires $O(m)$ storage before the caches are inserted.  The extra node-specific partial match queues add $O(m h)$ storage, since $\le m$ partial match queue entries are added to $\le h$ node-specific queues each.  The extra truncation refcount caches add $O(m k)$ storage, since we add $\le m k$ search window items to one sparse list each (sparse lists require linear storage).
\item
The lists of horizontal jumps take $O(k)$ storage, containing $\le k$ jumps each.
\end{itemize}

\section{Further Work}

It should be possible to find further optimizations for random text.

There are natural generalizations of copy-paste metrics to the graph matching problem, where pieces of one graph are reassembled to form another, with each piece and each ``glue'' edge or vertex imposing a cost.

\bibliographystyle{amsplain}
\bibliography{document}

\end{document}